\documentclass[preprint,nofootinbib,amsmath,prd,aps,superscriptaddress,tightenlines,12pt]{revtex4}
\usepackage[hypertex]{hyperref}
\usepackage{graphicx}
\usepackage{epstopdf}
\usepackage{bm}
\usepackage{epsfig}
\usepackage{graphics}
\usepackage{xspace}
\usepackage{subfigure}

\def\OMIT#1{}

\newcommand{\nn}{\nonumber}

\newcommand{\bea}{\begin{eqnarray}}
\newcommand{\eea}{\end{eqnarray}}

\newcommand{\gsim}{\mathrel{\rlap{\lower4pt\hbox{\hskip1pt$\sim$}}\raise1pt\hbox{$>$}}}

\newcommand{\be}{\begin{equation}}
\newcommand{\ee}{\end{equation}}

\begin{document}
\setlength\baselineskip{17pt}


\title{\bf  Color-Octet-Electroweak-Doublet Scalars and the CDF Dijet Anomaly}


\author{Linda M. Carpenter}
\affiliation{University of California at Irvine, Irvine, CA, 92697}
\email[]{lcarpent@uci.edu}
\author{Sonny Mantry}
\affiliation{University of Wisconsin, Madison, WI, 53706}
\email[]{mantry147@gmail.com}



\newpage
\begin{abstract}
  \vspace*{0.3cm}
We study the phenomenology of color-octet scalars in the $(8,2)_{1/2}$ representation in the context
of the  $3.2\sigma$ excess, in the dijet invariant mass spectrum of the $W+jj$ final state, recently observed by the CDF collaboration. We consider the region of parameter space with a sizable mass splitting between the charged and  neutral color-octet scalars and consistent with electroweak precision data. We implement the principle of Minimal Flavor Violation (MFV) in order to suppress FCNC currents and reduce the number of free parameters. The excess in the $W+jj$ channel corresponds to the charged current decay of the heavier neutral octet scalar into its lighter charged partner which decays into the two jets. In the MFV scenario, the production of the neutral color-octet is dominated by gluon fusion due to the Yukawa suppression of production via  initial state quarks. As a result,
 no visible excess is expected in the $\gamma+ jj$ channel due to Yukawa and CKM suppression.  Contributions to the $Z+jj$ final state are suppressed for a mass spectrum where the decay of the heavier color-octet to this final state is mediated by an off-shell neutral color-octet partner. MFV allows one to control fraction of bottom quarks in the final state jets by a single ratio of two free parameters.

\end{abstract}

\maketitle

\newpage
\section{Introduction}

New physics beyond the Standard Model (SM), that is likely to be discovered at the Tevatron or the Large Hadron Collider (LHC), is required to satisfy stringent constraints such as those arising from Electroweak Precision Data (EWPD)  or the smallness of Flavor Changing Neutral Currents (FCNC). In phenomenological analyses of extensions of the SM, compatibility with such constraints is facilitated by incorporating approximate symmetries such as custodial symmetry~\cite{Susskind:1978ms, Weinberg:1975gm, Sikivie:1980hm} or the principle of Minimal Flavor Violation (MFV)~\cite{Chivukula:1987py,Hall:1990ac, D'Ambrosio:2002ex}.  Recently~\cite{Arnold:2009ay}, it was shown that the possible extensions of the scalar sector of the SM are greatly restricted by the principle of MFV. In particular, any additional scalar degrees of freedom must have the quantum numbers $(1,2)_{1/2}$ or $(8,2)_{1/2}$ under the SM gauge group $SU(3)_C\times SU(2)_L\times U_Y(1)$.

In this work, we consider the extension of the  SM  by one family of color-octet scalars in the $(8,2)_{1/2}$ representation as given by the Manohar-Wise model~\cite{Manohar:2006ga}.  The phenomenology of color-octet scalars has been studied extensively in the literature~\cite{Gerbush:2007fe, Gresham:2007ri,Burgess:2009wm,FileviezPerez:2008ib, Plehn:2008ae, Arnold:2009ay, Han:2010rf, Dorsner:2009mq,FileviezPerez:2010ch, Idilbi:2009cc, Fornal:2010ac, Boughezal:2010ry,Boughezal:2011mh}. We revisit this scenario in the context of the recent  3.2$\sigma$   CDF anomaly~\cite{Aaltonen:2011mk}  in the dijet invariant mass ($M_{jj}$) distribution for the process $p\bar{p} \to W+jj$.  In particular, an excess was observed in the range $120 \>\text{GeV}\><M_{jj} <160\>\text{GeV}$. This excess is consistent with a dijet resonance with mass $144\pm 5$ GeV~\cite{Aaltonen:2011mk}. Several explanations for the observed excess have been  proposed using physics beyond the SM \cite{Eichten:2011sh,Nelson:2011us, Kilic:2011sr,Yu:2011cw, Wang:2011uq, He:2011ss, Rajaraman:2011rw, AguilarSaavedra:2011zy,Cheung:2011zt, Sato:2011ui, Wang:2011ta, Anchordoqui:2011ag, Dobrescu:2011px, Zhu:2011ww, Buckley:2011vs, Ko:2011ns, Fox:2011qd, Jung:2011ue, Cao:2011yt, Chang:2011wj, Harnik:2011mv, Hewett:2011nb,Fan:2011vw,Evans:2011wj, Gunion:2011bx, Faraggi:2011zi,Cheung:2011vx} and within the SM\cite{Plehn:2011nx, Sullivan:2011hu}. 

A more recent analysis by the  CDF collaboration~\cite{Annovi:2011aa}  found that the significance of the anomaly increased to the 4.1$\sigma$ level. This is consistent with expectations corresponding to the increase in the size of the data set from 4.3 fb$^{-1}$\cite{Aaltonen:2011mk} to 7.3 fb$^{-1}$\cite{Annovi:2011aa}. After the completion of this work, the results of the analysis by the D0 collaboration~\cite{Abazov:2011af} were released and no evidence of a new resonance in the $M_{jj}$ distribution was found. We await the results of the analysis of the joint task force~\cite{Eichten:2011aa} aimed at understanding the discrepancy between the CDF and D0 results. 

\begin{figure}
\includegraphics[height=2.5in, width=3in]{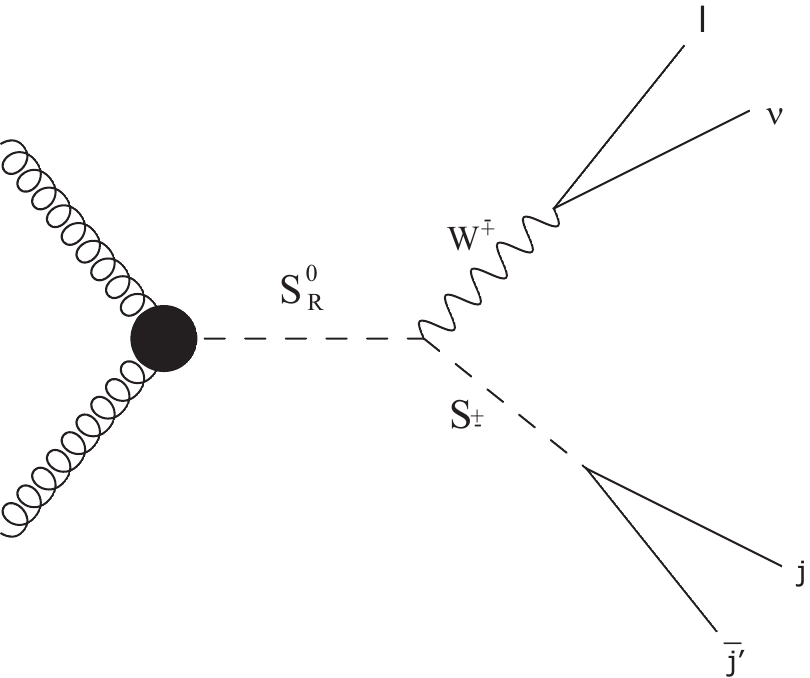}
\caption{Production of the color-octet neutral states $S_{R,I}^0$ via gluon fusion followed by the on-shell decays $S_{R,I}^0\to S^\pm W^\mp$, $S^\pm \to jj$ and $W^\pm \to \ell \nu$. }
\label{SR-WSpm}
\end{figure}

The $(8,2)_{1/2}$ color-octet-electroweak-doublet of the Manohar-Wise model consists of four states: $S_R^0$, $S_I^0$, and $S^\pm$. The states $S_R^0$ and $S_I^0$ correspond to the real and pseudoscalar neutral components of the doublet respectively. The $S^\pm$ states correspond to the charged components of the doublet. The masses of the states $S_R^0,S_I^0$, and $S^\pm$ are denoted by $M_R,M_I$, and $M_\pm$ respectively. For significant mass splittings between the charged and neutral states, the process (see Fig.~\ref{SR-WSpm})
\bea
\label{process}
p\bar{p}\to S_{R}^{0}\to W^\pm S^\mp \to \ell \bar{\nu}_\ell + jj,
\eea
allows for an explanation of the CDF anomaly. The observed excess in the $M_{jj}$ spectrum would correspond to the on-shell decay $S^\pm\to jj$ for $M_\pm \sim 144$ GeV. A mass splitting of $M_R- M_\pm \geq 80$ GeV is required for  $S_R^0$ to also be on-shell in Eq.(\ref{process}). Such a large mass splitting has been shown to be consistent with EWPD~\cite{Burgess:2009wm}. Future analyses can distinguish such an s-channel production mechanism from models with t-channel production by looking for the $S_R^0$ resonance in the invariant mass distribution  of the $Wjj$ system~\cite{Eichten:2011sh}.

We employ the principle of MFV to avoid dangerous tree-level FCNCs. Implementing MFV also dramatically reduces  the number of free parameters allowing for a more predictive framework. In particular, the coupling of the color-octet scalars to the SM fermions is determined entirely in terms of the SM Yukawa matrices  $g^{U,D}$ and just two unknown  proportionality constants $\eta_{U,D}$ respectively. In addition to the structure of the Yukawa matrices, the overall strength of the coupling of the color-octet scalars  to the SM fermions can be controlled by the size of the free parameters $\eta_{U,D}$. The region of very small $\eta_{U,D}$ would correspond to a fermiophobic octet scalar. Such a fermiophobic  color-octet scalar is also a viable option since it can still be produced via gluon fusion involving color-octet scalar loops.

\begin{figure}
\includegraphics[height=2in,width=3in]{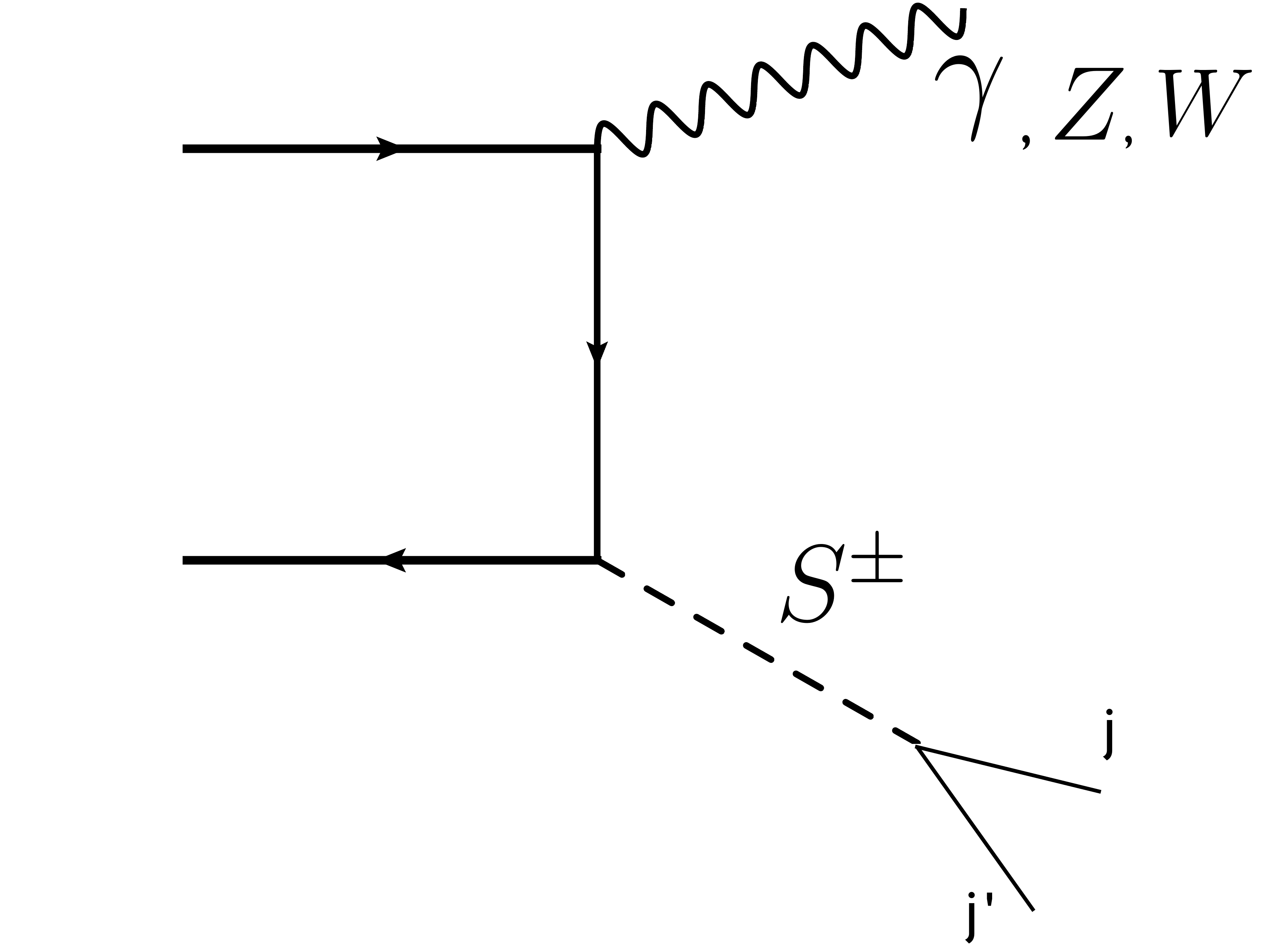}
\caption{Production of charged color-octet scalars from initial state quarks in association with a $\gamma, Z,$ or $W$. This production channel suppressed by a Yukawa factor at the $S^\pm$ production vertex in MFV.}
\label{Zjj}
\end{figure}

Below we summarize the relevant features of the Manohar-Wise model:
\begin{itemize}
\item The spectrum of color octet states required to explain the CDF anomaly is as follows. The charged states have mass $M_\pm \sim 144$ GeV and correspond to the resonance seen in the $M_{jj}$ distribution. The $S_R^0$ state is required to be much heavier with $M_R-M_\pm \geq 80$ GeV to allow for the on-shell decay $S_R^0\to WS^\pm$. Consistency with EWPD requires $|M_I -M_\pm| < 50$ GeV. For the numerical analysis and discussion in most of the paper, we use the benchmark point $M_R=245$ GeV, $M_I=190$ GeV, and $M_\pm =150$ GeV with a   cross-section for the production of $S_R^0$ of about 1.8 pb to explain the CDF anomaly.

\item Due to the principle of MFV there are  two  parameters, $\eta_{U}$ and $\eta_D$,  determine the couplings of the color-octet scalars to the SM fermions.  Kinematically, the charged states $S^\pm$ can only decay into $jj$ involving the  quark flavors $u,d,c,s,b$. The precise flavor structure of the jet pair $jj$ is determined by the SM
Yukawa couplings and the CKM matrix. For example, for $\eta_U \gg \eta_D$, the state $S^+$ will dominantly decay to $c\bar{s}$ due to the relative Yukawa or CKM suppression of the other channels. One  has the flexibility to increase the fraction of b-quarks in $jj$ by choosing  larger values for the ratio $\eta_D/\eta_U$.

\item The s-channel production of $S_R^0$ will be dominated by gluon fusion due to the Yukawa suppression of production via $q\bar{q}$ annihilation. Gluon fusion production can proceed via a top quark or  color-octet scalar loop. The size of contribution of the top quark loop can be controlled via the free parameter $\eta_U$. 

\item The additional contribution of the Manohar-Wise model to the process $p\bar{p}\to \gamma + jj$ is Yukawa suppressed.  The contribution is shown in Fig.~\ref{Zjj} where the $S^\pm jj$ vertex is proportional to a Yukawa coupling and a CKM factor. As a result, the contribution to the $M_{jj}$ spectrum in the $\gamma +jj$ channel is negligible compared to background processes. Even with improved future experimental analyses,  no excess should be visible in the $M_{jj}$ distribution of the $\gamma +jj$ channel. This mechanism and degree of suppression in the $\gamma+jj$ channel is unique feature of the Manohar-Wise model. 

\item For the benchmark spectrum $M_R=245$ GeV, $M_I=190$ GeV, and $M_\pm =150$ GeV, the contribution to the $p\bar{p}\to Z + jj$ channel will be suppressed relative to background processes. This suppression occurs because the decay $S_R^0\to Z+jj$ is mediated by an off-shell $S_I^0$ state via $S_R^0 \to ZS_I^{0*}\to Z+jj$. If one chooses a mass spectrum where $M_R-M_I\geq 90$ GeV, the production rate for the $Z+jj$ state will be comparable to that of the $W+jj$ state. If an excess is observed in the $Z+jj$ channel in future analyses, such a mass spectrum with $M_R-M_I\geq 90$ GeV will be worth further study .

\end{itemize}

The outline of the paper is as follows. In section \ref{octet-model} we review the scalar color-octet Manohar-Wise model. In section \ref{precision} we discuss the region of parameter space suited for explaining the CDF $Wjj$ anomaly as well as the implications of precision constraints. In section\ref{decay-prod} we discuss the production and decay properties of the color-octet scalars. In section \ref{simulation} we show simulation results and give conclusions in section~\ref{conclusions}.

\section{Color Octet Scalar Model}
\label{octet-model}

Here we review the basics of the Manohar-Wise model~\cite{Manohar:2006ga} for a new color-octet coupled to the SM with gauge quantum numbers $(8,2)_{1/2}$ under the SM gauge group $SU(3)_C\times SU(2)_L\times U(1)_Y$. The $(8,2)_{1/2}$ color octet state $S^A$ is given by
\bea
(S^A)^T = \big (S^{A+} , S^{A0} \big ),
\eea
where $A$ denotes the color index and $S^{A+}$ and $S^{A0}$ charged and neutral electroweak doublet components and
\bea
S^{A0} &=& \frac{S_R^{A0}+ i S_I^{A0}}{\sqrt{2}},
\eea
where $S_R^{A0}$ and $S_I^{A0}$ denote the real scalar and pseudoscalar neutral components. The couplings
to the SM fermions in accordance with the principle of MFV are given by
\bea
\label{yukawa}
{\cal L} = -\eta_{U} g_{ij}^U \bar{u}_R^i T^A (S^A)^T \epsilon Q_L^i - \eta_D g_{ij}^D \bar{d}_R^i T^A (S^A)^\dagger Q_L^j + h.c,
\eea
where $g_{ij}^{U,D}$ are the SM Yukawa matrices, $i,j$ are flavor indices and $\eta_{U,D}$ are overall constants. For simplicity we assume that $\eta_{U,D}$ are real and there is no mixing between the $S_R^0$ and the $S_I^0$ states.
The most general renormalizable scalar potential in this model is given by~\cite{Manohar:2006ga}
\bea
V&=& \frac{\lambda}{4} \big (H^{\dagger i} H_i - \frac{v^2}{2}\big )^2 + 2m_S^2 \text{Tr} (S^{\dagger i}S_i) + \lambda_1 H^{\dagger i}H_i \text{Tr} (S^{\dagger j}S_j) + \lambda_2 H^{\dagger i}H_j \text{Tr} (S^{\dagger j}S_i) \nn \\
&+& \big [ \lambda_3 H^{\dagger i} H^{\dagger j} \text{Tr}(S_iS_j) + \lambda_4 H^{\dagger i} \text{Tr}(S^{\dagger j}S_j S_i) +  \lambda_5 H^{\dagger i} \text{Tr}(S^{\dagger j}S_i S_j) + h.c \big ] \nn \\
&+& \lambda_6 \text{Tr} (S^{\dagger i}S_i S^{\dagger j} S_j) + \lambda_7 \text{Tr}(S^{\dagger i}S_j S^{\dagger j} S_i) + \lambda_8 \text{Tr} (S^{\dagger i}S_i)\text{Tr}(S^{\dagger j}S_j) \nn \\
&+& \lambda_9 \text{Tr} (S^{\dagger i} S_j) \text{Tr}(S^{\dagger j}S_i) + \lambda_{10} \text{Tr} (S_i S_j) \text{Tr}(S^{\dagger i}S^{\dagger j}) +\lambda_{11} \text{Tr}(S_iS_j S^{\dagger j} S^{\dagger i}),
\eea
where the $i,j$ denote $SU(2)_L$ indices and $S=S^AT^A$. After electroweak symmetry breaking, the mass spectrum of the charged and neutral components of the color-octet scalars depend on the parameters in the scalar potential and at tree-level are given by
\bea
\label{mass}
M_\pm^2 &=& m_S^2 + \lambda_1 \frac{v^2}{4}\nn \\
M_R^2 &=& m_S^2 + (\lambda_1 + \lambda_2 + 2\lambda_3) \frac{v^2}{4} \nn \\
M_I^2 &=& m_S^2 +(\lambda_1 + \lambda_2 - 2\lambda_3) \frac{v^2}{4}.
\eea

\section{Precision Constraints}
\label{precision}

As discussed in the introduction,  the CDF anomaly can be explained via the production of $S_{R}^0$ followed by the on-shell  cascade decays $S_{R}^0\to W^\pm S^\mp$, $W\to \ell \nu$, and $S^\pm \to jj$:
\bea
p\bar{p} \to S_{R}^0 \to W^\pm S^\mp \to \ell \nu + jj.
\eea
The excess in the dijet invariant mass spectrum would correspond to the on-shell decay $S^\pm \to jj$ for $M_\pm \sim 144$ GeV. We allow for a large enough mass splitting between $M_{R}$ and $M_\pm$ to allow for the on-shell charged current decay $S_{R}^0\to W^\pm S^\mp$. At the same time, we choose the mass splitting between $M_R$ and $M_I$ to be not too large so that the on-shell decay $S_{R}^0\to Z^0S_{I}^0$, which could contribute in the $Z+ jj$ channel, is kinematically forbidden.

In Ref.~\cite{Burgess:2009wm}, an in depth analysis was performed to determine the regions of parameter space in the color-octet model that are consistent with EWPD, collider constraints, and FCNC constraints. Based on their analysis, there are regions in parameter space where the mass spectrum of interest in consistent with EWPD. The most robust constraints come from direct searches for pair production of octet scalars at LEP requiring
\bea
M_\pm > 100 \>\text{GeV}, \qquad M_R + M_I > 200 \> \text{GeV}.
\eea
The octet scalars affect electroweak precision observables through their contributions to the oblique corrections. These constraints were examined in detail in Ref.~\cite{Burgess:2009wm} and it was found that the strongest correlation was between the masses $M_I$ and $M_\pm$ requiring $|M_I-M_\pm|< 50$ GeV in the 95\% confidence region. This correlation can be traced to a custodial symmetry which, when exact, requires~\cite{Manohar:2006ga, Burgess:2009wm}
\bea
2\lambda_3 &=&\lambda_2, \qquad 2\lambda_6 = 2\lambda_7 = \lambda_{11}, \nn \\
\lambda_9 &=& \lambda_{10}, \qquad \lambda_4=\lambda_5^*
\eea
As seen from Eq. (\ref{mass}), in the limit of exact custodial symmetry, the condition $2\lambda_3 = \lambda_2$ is equivalent to $M_\pm=M_I$. Since EWPD favors an approximate custodial symmetry, there is a strong correlation between $M_\pm$ and $M_I$ in EWPD fits. We note that the EWPD constraint $|M_I-M_\pm|< 50$ GeV, naturally disfavors the on-shell decay channel $S_{I}^0\to S^\pm W^\mp$. For this reason, we only consider the single production of $S_R^0$, and not $S_I^0$, as a means to explain the CDF anomaly.

\begin{figure}[h]
\begin{center}$
\begin{array}{ccc}
\includegraphics[height=3.0in, width=3.0in]{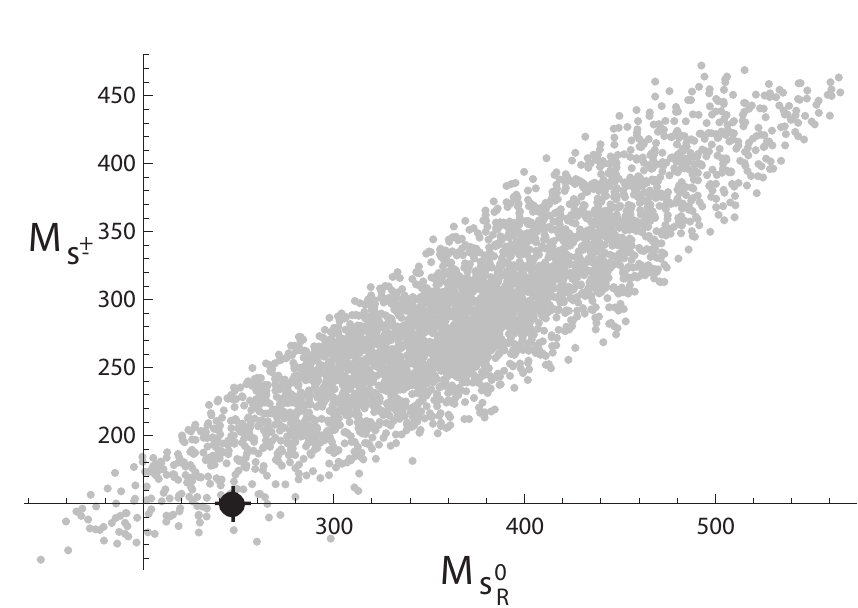} &
\includegraphics[height=3.0in,width=3.0in]{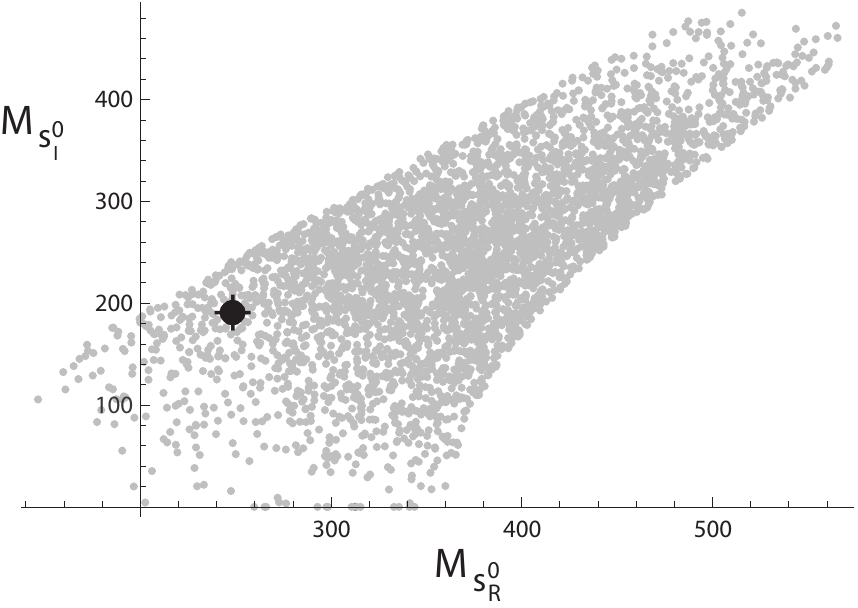} 
\end{array}$
\end{center}
\caption{Allowed regions in the $S_R^{0}$,$S^\pm$ and
$S_I^{0}$,$S_R^{0}$ mass planes which are consistent with contributions to
the S and T parameters at $2\sigma$. The benchmark point of $M_R=245$ GeV, $M_I=190$ GeV, and $M_\pm=150$ GeV, which lies in the 2$\sigma$ allowed region, is identified as a small  black shaded circle for convenience. For a more detailed analysis see Ref.~\cite{Burgess:2009wm}}
\label{EW}
\end{figure}

For purposes of illustration, in  Fig.~(\ref{EW}) we plot the allowed regions in the $S_R^{0}-S^\pm$ and
$S_I^{0}-S_R^{0}$ mass planes which are consistent with contributions to
the S and T parameters at $2\sigma$. Notice that though both the masses of
$S^\pm$ and $S_I^{0}$ cannot differ from $S_R^{0}$ by hundreds of GeV, our
region of intermediate mass splitting of order 100 GeV is not disfavored. For convenience, in Fig.~(\ref{EW}) we have identified the benchmark point of $M_R=245$ GeV, $M_I=190$ GeV, and $M_\pm=150$ GeV used in the numerical analysis. For a more detailed and extensive analysis see Ref.~\cite{Burgess:2009wm}.

Implementing the principle of MFV to determine the couplings of the octet scalars to the SM fermions, avoids dangerous new tree-level FCNC processes. However, beyond tree level, FCNC can be generated giving rise to constraints on the MFV parameters $\eta_{D,U}$ in Eq. (\ref{yukawa}). These constraints were examined in ~\cite{Manohar:2006ga,Gresham:2007ri,Burgess:2009wm}. In particular, contributions to $K^0-\bar{K}^0$ mixing, $B\to s\gamma$, $R_b$, non-leptonic B-decay decays via $b\to sg$, and the neutron electric dipole moments were studied. In our case, we find a sizable cross-section for the production of $S_{R}^0$ via the color octet scalar loops  (last two diagrams in Fig.~\ref{prod}) as long as $(\lambda_4 +\lambda_5)$ is not too small. This allows us to choose $\eta_{U,D}$ to be small enough to avoid precision constraints. 

\section{Production and Decay}
\label{decay-prod}

\begin{figure}[h]
\begin{center}$
\begin{array}{ccc}
\includegraphics[height=1.5in, width=2.0in]{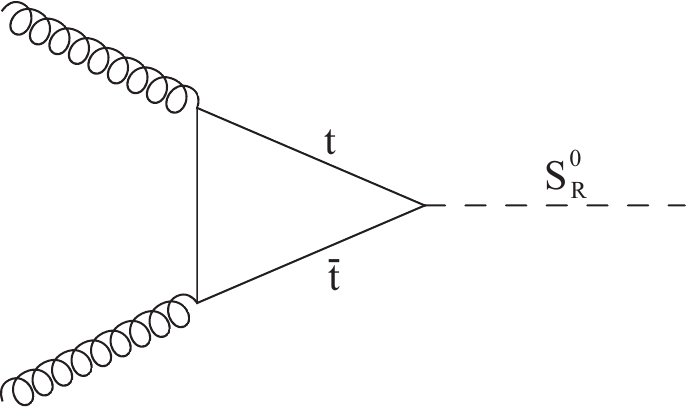} &
\includegraphics[height=1.5in,width=2.0in]{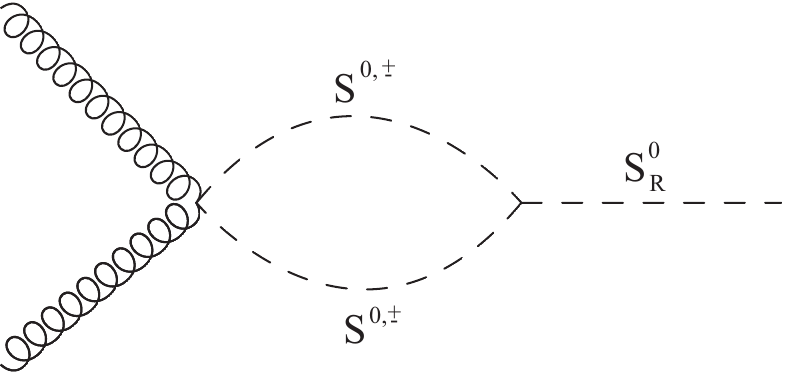} &
\includegraphics[height=1.5in,width=2.0in]{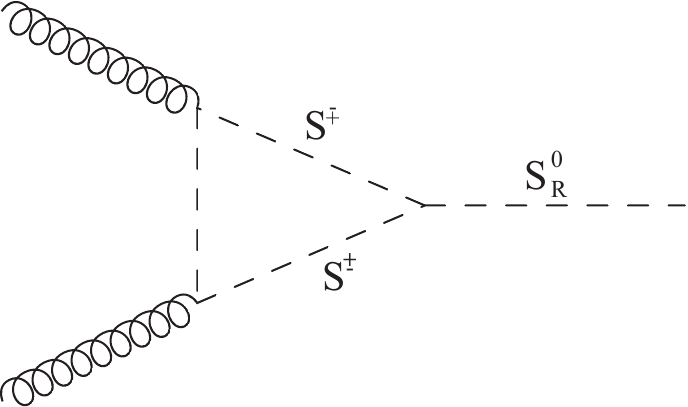}
\end{array}$
\end{center}
\caption{Production of $S_R^0$ via gluon fusion.}
\label{prod}
\end{figure}

The production of the  neutral scalars $S_{R,I}^0$ will be dominated by gluon fusion as shown in Fig.~\ref{prod}.  In this case, the production cross-section~\cite{Gresham:2007ri}
can be related to the decay rate as
\bea
\label{production}
\sigma (p\bar{p} \to S^0_{R,I}) &=& \frac{1}{16}\Gamma(S_{R,I}^0\to gg) \frac{16 \pi^2}{s M_R} \>\int_{m_S^2/s}^1 \frac{dx}{x} g(x) g(\frac{M_R^2}{s\> x}),
\eea
where the factor of $1/16$ appears in relating the decay rate to the production cross-section. While we are only interested in the production of $S_R^0$ in this paper, we also give the expression for the production of $S_I^0$ in order to facilitate discussion later in this section. The expression for the decay rate $\Gamma (S_R^0\to gg)$, appearing in Eq.(\ref{prod}),  was given in ~\cite{Gresham:2007ri} for the case of a degenerate color-octet mass spectrum. For the case of arbitrary mass splittings between the electroweak components of the octet scalars we find
\bea
\label{decay}
\Gamma (S_R^0\to gg) &=& \frac{G_F M_R^3\alpha_s^2}{\sqrt{2} 2^{10}\pi^3} \Bigg \{ \frac{C_1 \eta_U^2}{9} |A_t(\tau_t)|^2 + \frac{4C_3}{9} (\lambda_4+\lambda_5)^2 \frac{v^4}{M_R^4}\Big | \sum_i \frac{f_i}{\tau_i} A_{S}(\tau_i)\Big |^2   \nn \\
&+& \frac{4C_2}{9} (\lambda_4+\lambda_5)\eta_U \frac{v^2}{M_R^2}\text{Re} \Big [ A_t(\tau_t) \sum_i \frac{f_i}{\tau_i} A_S(\tau_i)\Big ]\Bigg \},
\eea
where for simplicity we have assumed that $\lambda_{4,5}$ and $\eta_U$ are real. The index $i$ in $f_i, \tau_i$ runs over the states $S^{\pm},S_R^0$, and $S_I^0$ that run in the loops in Fig.~\ref{prod}. The $f_i$ take on the values
\bea
f_\pm &=& 1, \qquad f_R= 3, \qquad f_I = 1,
\eea
and $\tau_t$ and $\tau_i$ are defined as
\bea
\tau_t &=& \frac{4m_t^2}{M_R^2}, \qquad \tau_i = \frac{4M_i^2}{M_R^2}.
\eea
The color factors $C_{1,2,3}$ are defined as
\bea
C_1 = \sum (d^{ABC})^2, \qquad C_2 =\sum_{A,B,C} d^{ABC}d^{GFC}f^{AEF}f^{BGE}, \qquad C_3 = \sum (d^{GFC}f^{AEF}f^{BGE})^2, \nn \\
\eea
and take on the numerical values $C_1=40/3$, $C_2=-20$, and $C_3=30$.
$A_t(\tau)$ and $A_S(\tau)$ are the standard functions~\cite{Djouadi:1991tka, Dawson:1990zj, Graudenz:1992pv, Kauffman:1993nv, Dawson:1993qf, Spira:1997dg,Djouadi:2005gi,Djouadi:2005gj,Spira:1995rr} that appear in the phenomenology of
Higgs decays in the MSSM and are given by
\bea
A_t(\tau) &=& \frac{3}{2} \tau [1+(1-\tau) f(\tau)], \qquad A_S(\tau) = -\frac{3}{4}\tau [1-\tau f(\tau)],
\eea
where the function $f(\tau)$ is defined as
\bea
f(\tau)=
\begin{cases}
\text{arcsin}^2 \frac{1}{\sqrt{\tau}}, &\tau \geq 1\\
-\frac{1}{4}\Big [ \ln \frac{1+ \sqrt{1-\tau}}{1-\sqrt{1-\tau}}-i\pi \Big ]^2, &\tau <1.
\end{cases}
\eea
\begin{figure}[h]
\includegraphics[height=2.5in, width=6.0in]{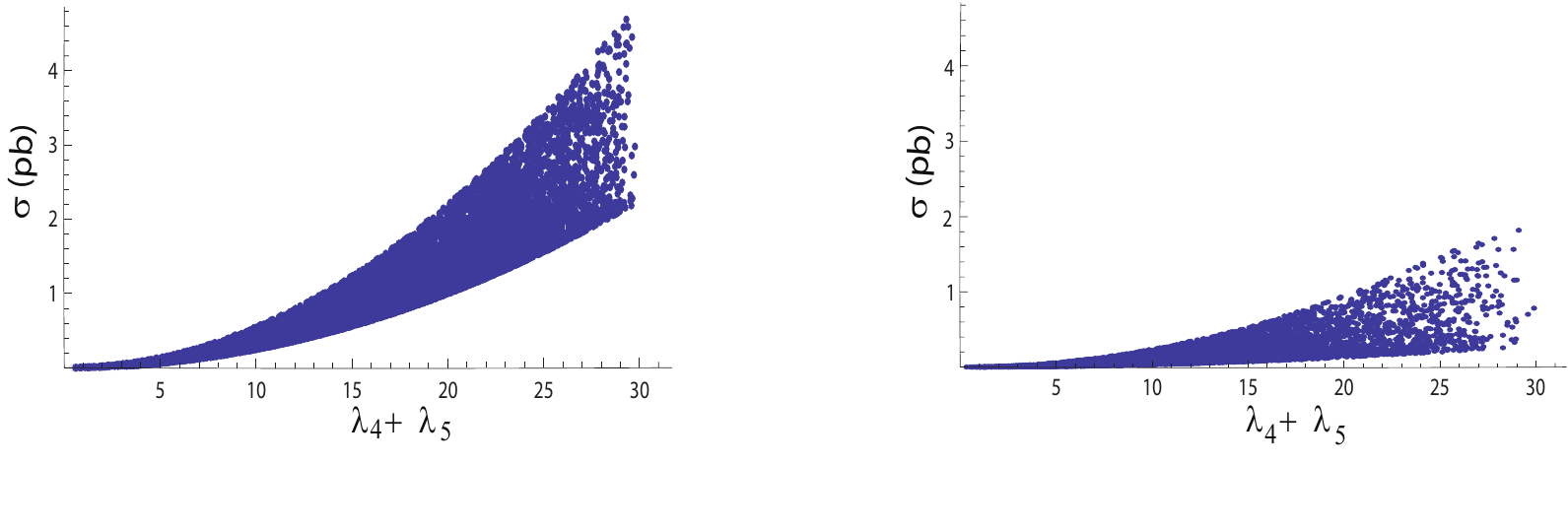} 
\caption{Production cross-section of $S_R^0$ as a function of $(\lambda_4 + \lambda_5)$. The plot on the left is for $M_\pm =150$ GeV for the ranges $230\> \text{GeV} < M_R < 300\text {GeV}$ and $150 \text{GeV} < M_I < 200 GeV$. The plot on the right corresponds to a more degenerate spectrum where $230\> \text{GeV}< M_R < 300$ GeV and $M_\pm, M_I$ are within 10 GeV of $M_R$. The CDF anomaly requires a non-degenerate mass spectrum contained in the region of the left panel above for which the cross-section is enhanced compared to the case of a degenerate mass spectrum (right panel). We have set $\eta_U=0$ so that the contribution to the cross-section from top quark loops is not included in the above plots.}
\label{prod-deg}
\end{figure}
Note that the production cross-section for $S_R^0$ given by Eqs.(\ref{production}) and (\ref{decay}) depends on the parameters $\eta_U$ and $(\lambda_4+\lambda_5)$. In particular, in the limit that $\eta_U$ is taken to be very small so that the top loop contribution is turned off, one can still generate a sizable cross-section through the contribution of the octet-scalar
loops through the coupling $(\lambda_4 + \lambda_5)$. We also note that for a degenerate mass spectrum for the color-octet scalars, the production cross-section for $S_R^0$ is suppressed~\cite{Gresham:2007ri} by a numerical factor of $(\pi^2/9-1)^2$. However, for significant mass splittings between the color-octet states, this numerical suppression is lifted and the cross-section is enhanced. In Fig.~\ref{prod-deg} we compare the production cross-section of $S_R^0$ for the case of a non-degenerate color-octet mass spectrum (left panel) with that for a degenerate
cross-section (right panel). For simplicity we have set $\eta_U=0$ so that the additional contribution to the cross-section from top quark loops (first diagram in Fig.~\ref{prod}) is not included. We see that the cross-section is relatively enhanced for a non-degenerate spectrum (left panel). An explanation of the CDF anomaly requires a non-degenerate mass spectrum so that the relevant production cross-section is obtained from the left panel of Fig.(\ref{prod-deg}). In the left panel, we have also allowed for points where the decay $S_R^0\to Z+jj$ can proceed via an on-shell intermediate $S_I^0$ state in order to give typical cross-sections over a wider range of parameter space. 
  As seen from the left panel of Fig.(\ref{prod-deg}),
a large enough cross-section ($\sim$ 1.8 to 2 pb) needed to explain the CDF anomaly is possible for large but still perturbative values of $\lambda_4$ and $\lambda_5$. Relatively large values of $\lambda_4$ and $\lambda_5$ are needed to overcome the loop suppression for production via gluon fusion. For the benchmark point of $M_R=245$ GeV, $M_I=190$ GeV, and $M_\pm=150$ GeV, we find that $\lambda_4+\lambda_5=18$ is sufficient to explain the CDF anomaly and gives a cross-section of 1.8 pb. This can be obtained by choosing $\lambda_4\sim \lambda_5 \sim 9$ which is still in the perturbative range ($\lambda_{4,5}\lesssim 10$)~\cite{Burgess:2009wm}. 

While the single production of $S_I^0$ is not of direct interest to explaining the $W+jj$ excess, due to the EWPD data preference~\cite{Burgess:2009wm} of $|M_I-M_\pm |<50$ GeV, we still briefly comment on its properties. The production of $S_I^0$ occurs only via the top quark loop as shown in the first diagram in Fig.~\ref{prod} for real values of $\lambda_{4,5}$. In this case, the expression for its decay rate to two gluons is given by~\cite{Gresham:2007ri}
\bea
\label{decay2}
\Gamma (S_I^0 \to gg) &=& \frac{G_F\alpha_s^2 m_t^4}{M_I\sqrt{12}\pi^3}C_1|\eta_U|^2 | 4f(\tau_t)|^2.
\eea
Eqs.(\ref{production})  and (\ref{decay2}) together give the production rate for $S_{I}^0$. Note that unlike $S_R^0$, one can turn off the production of $S_I^0$ by choosing smaller and smaller values for $\eta_U$ when $\lambda_{4,5}$ are real.  Finally, charge conservation does not allow the single production of $S^\pm$ via gluon fusion. Instead the production of $S^\pm$ will proceed through  $q\bar{q}'$ annihilation and is independent of $\lambda_{4,5}$. Furthermore,  the production of $S^\pm$ via $q\bar{q}'$ annihilation will always be Yukawa and CKM suppressed due to  MFV. 

At tree level, the neutral color-octet scalar $S_R^0$ can decay as $S_R^0\to S^\pm W^\mp$ and $S_R^0\to q\bar{q}$. The size of the branching fraction $\text{Br}(S_R^0\to S^\pm W^\mp)$, the channel of interest to explain the excess in the $W+jj$, relative to $\text{Br}(S_R^0\to q\bar{q})$ can be enhanced by choosing smaller values of $\eta_{U,D}$ in the MFV scenario. The decay rate $\Gamma(S_R^{0}\to S^{\pm} W^\mp)$ is given by
\bea
\Gamma (S_R^{0}\to W^\pm S^{\mp})  &=& \frac{g^2}{64\pi (N_c^2-1)M_R^3}\Big [ \Big ( M_R^2 - (M_\pm + M_W)^2\Big )\Big ( M_R^2-(M_W-M_\pm)^2\Big )\Big ]^{1/2}\nn \\
&\times&  \Big [ 2M_R^2 + 2M_\pm^2 -M_W^2 - \frac{(M_R^2-M_\pm^2)^2}{M_W^2}\Big ],
\eea
and the decay rates to fermions can be found in \cite{Manohar:2006ga}. Additional one-loop-supprsessed decay channels $S_R^0\to gg, \gamma \gamma, WW, ZZ$ are also available.

For the mass spectrum where $M_\pm < M_{R,I}$ and for $\eta_D \ll \eta_U$, the charged states $S^\pm$ will dominantly decay as $S^+\to c\bar{s}$. The decays $S^+\to u(\bar{d},\bar{s},\bar{b})$ will be Yukawa suppressed in MFV due to the small up quark mass and the decays to top quarks $S^+\to t(\bar{d},\bar{s},\bar{b})$ are kinematically forbidden for $M_\pm \sim 144$ GeV. The decays $S^+\to c(\bar{d},\bar{b})$ will be CKM suppressed compared to $S^+\to c\bar{s}$. As a result, the dominant branching fraction will be for $S^+\to c\bar{s}$.

Kinematically, the only allowed decay channel for $S^\pm$ is $S^\pm \to jj$ so that choosing smaller values of $\eta_{D,U}$ will not affect the branching fraction $\text{Br}(S^\pm \to jj)$. One can change the relative branching fractions for $S^\pm$ decays into different quark flavors, by changing the relative sizes of $\eta_U$ and $\eta_D$. For example, one can choose both $\eta_{U}$ and $\eta_D$ small enough to avoid precision constraints, but still impose the hierarchy $\eta_U\gg \eta_D$ to suppress decays involving a bottom quark  or $\eta_U\ll \eta_D$ to enhance such decay channels.

\section{Simulation}
\label{simulation}
Here we show simulation results for the process $p\bar{p}\to S_R^0\to  W^\mp S^\pm \to \ell \nu + jj$. We have generated Events using MADGRAPH, which were showered with PYTHIA and run through the detector simulator PGS.  Cuts have been implemented matching those of the CDF analysis.  Events were required to have exactly two jets, of cone size $R=0.4$ with  $E_T > 30$ GeV, and $|\eta| <2.4$.  Cuts were placed on the dijet angles requiring $|\eta_1- \eta_2| <2.5$ and $|\delta \phi|< 0.4$. Events were required to have more than 25 GeV of missing energy, and an isolated lepton was required with $p_T > 20$ GeV, with $M_T$ of the missing energy and lepton to be $> 30$ GeV.  An additional cut rejected jets within $R< 0.52$ of the isolated lepton, and an additional cut was placed to veto more than one isolated lepton with $p_T>10$ GeV.

\begin{figure}[h]
\centerline{\includegraphics[width=10 cm]{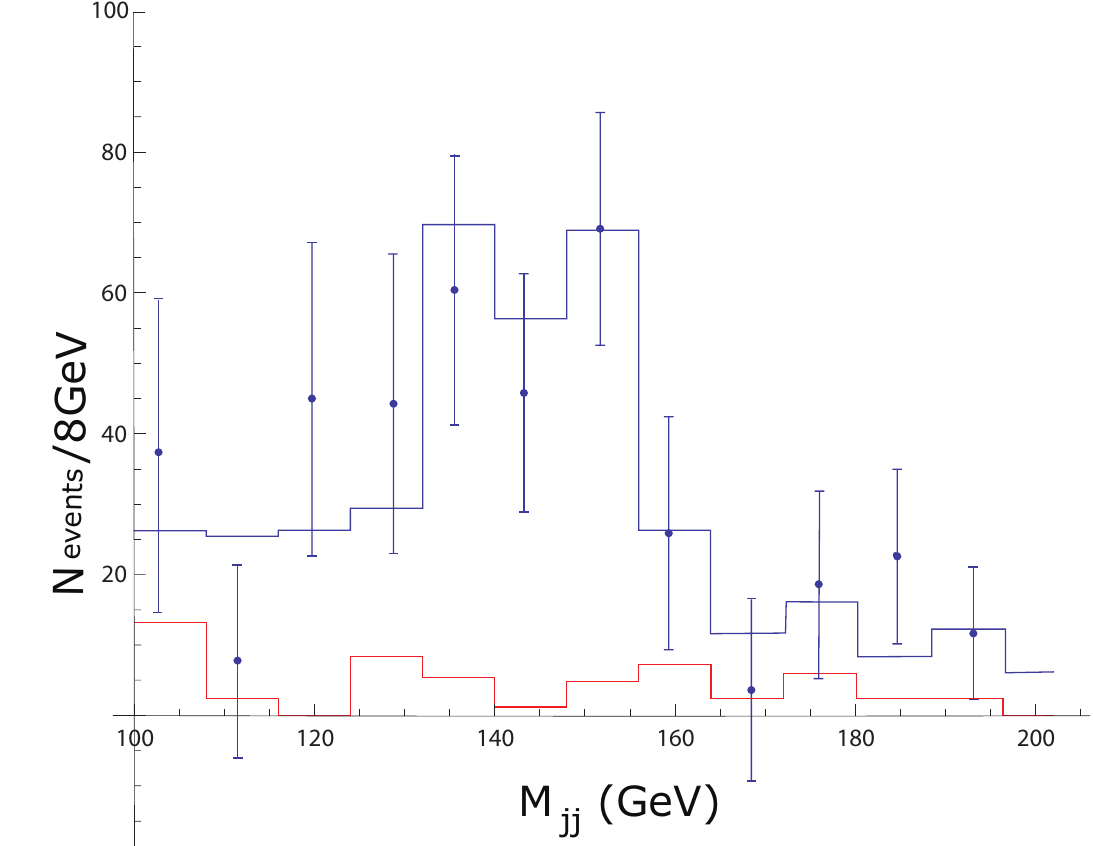}}
\caption{Plot of N events is 4.3 fb$^{-1}$ of Tevatron data vs. dijet invariant mass for  $M_{R} = 245$ GeV, $m_{S\pm}=150$ GeV,  $M_I=190$ GeV, and $\lambda_4+\lambda_5=18$. This corresponds to a production cross-section for $S_R^0$ of 1.8 pb. Also shown are Tevatron data points with error bars and the tail of the QCD subtracted dijet distribution from di-boson production (bottom red line).}
\label{fig:2pgm}
\end{figure}

Fig.~\ref{fig:2pgm} shows the number of expected events vs the dijet invariant mass for $M_R=245 $ GeV, $M_I=190$ GeV, $M_\pm = 150$ GeV and $(\lambda_4+\lambda_5)=18$. This requires large couplings $\lambda_4\sim \lambda_5\sim 9$ which are still in the perturbative range $\lambda_{4,5}\lesssim 10$~\cite{Burgess:2009wm}. The relatively large couplings are required to overcome the loop suppression for production via gluon fusion. The events are binned in 8 GeV steps. Agreement with data is good, as one can see with comparison to the Tevatron data points~\cite{Aaltonen:2011mk} after background subtraction, which has also been plotted.

\section{Conclusions}
\label{conclusions}

We have investigated the phenomenology of color-octet scalars in the $(8,2)_{1/2}$ representation of the SM gauge group $SU(3)_C\times SU(2)_L\times U(1)_Y$ in the context of the recently observed excess in the dijet invariant mass spectrum in the $p\bar{p}\to W +jj$ channel. We consider the region of parameter space where there is a significant mass splitting between the color-octet states $S_R^0,S_I^0,S^\pm$ while still remaining consistent with EWPD. We employed the principle of MFV to avoid new FCNCs and have a more predictive framework for the flavor structure of the theory. The couplings of the color-octet scalars are determined in terms of the Yukawa and CKM matrices and two unknown parameters $\eta_{U,D}$ that control the overall size of the couplings.

We find that there are regions in parameter space of the color-octet model that can explain the recently observed~\cite{Aaltonen:2011mk} $3.2\sigma$ excess in the dijet invariant mass spectrum in $p\bar{p}\to W+jj$ while simultaneously avoiding a corresponding excess in the $\gamma +jj$ and $Z+jj$ channels. Other models where excesses are predicted in the $\gamma +jj$ and $Z+jj$ channels are not necessarily ruled out~\cite{Jung:2011ua, Eichten:2011xd} by current data. If excesses are also seen in the $Z+jj$ channel, a different choice of mass spectra in the Manohar-Wise model can accommodate such a scenario. However, in the Manohar-Wise model, an excess in the $\gamma +jj$ channel cannot be produced due to Yukawa and CKM suppression mechanisms. The signal in the $W+jj$ channel can be explained via the process $p\bar{p}\to S_R^0\to W^\pm S^\mp\to \ell \nu + jj$ where excess in the dijet spectrum would correspond to $S^\pm \to jj$.   This requires a significant mass splitting between the $S_R^0$ and $S^\pm$  states to kinematically allow the decay $S_R^0\to S^\pm W^\mp$. We also choose a mass splitting between $S_R^0$ and $S_I^0$ that is not too large in order to kinematically forbid the decay $S_R^0\to S_I^0Z^0$ which would otherwise contribute an excess in the $Z+jj$ channel.  The production of $S_R^0$ is dominated by gluon fusion since the production quark-anti-quark annihilation channel is Yukawa suppressed in the MFV scenario. This automatically suppresses contributions to the $\gamma +jj$ final state which occurs via a photon radiated from the initial state quarks. Further prospects may include studies of octet pair production with cascade decay topologies yielding final states with multiple jets and gauge bosons. For example, pair production of $S_R^0$ could give final states with $W^+W^-S^+ S^-$ which can give a hard lepton and multiple jets. This could be interesting signature to study.

There are regions in the parameter space of the color-octet $(8,2)_{1/2}$ model that are consistent with EWPD and able to explain the $W+jj$ excess observed by the CDF collaboration. Combined with MFV this is can be a predictive framework that is worth further study.

\acknowledgments{We thank Gil Paz, Arvind Rajaraman, Tim Tait,  Maike Trenkel, and Michael Trott for useful discussions. This work was supported in part under U.S. Department of Energy contract DE-FG02-08ER4153, the Wisconsin Alumni Research Foundation, and the NSF Grant No. PHY-065365.}
\bibliographystyle{h-physrev3.bst}
\bibliography{octet}

\end{document}